\documentclass{article}
\usepackage{graphicx} %
\usepackage{fancyhdr}
\usepackage{xcolor}

\usepackage{xspace}
\usepackage{url}

\newcommand{\sakshi}{{\tt SAKSHI}\xspace}

\title{\sakshi: Decentralized AI Platforms}
\author{Suma Bhat$^{1,3}$\thanks{Authors are listed alphabetically.\\ Correspondance to : \{hebbar, pramodv\}@princeton.edu} \and Canhui Chen$^2$ \and Zerui Cheng$^2$ \and Zhixuan Fang$^2$ \and Ashwin Hebbar$^1$ \and Sreeram Kannan$^5$ \and Ranvir Rana$^4$ \and Peiyao Sheng$^3$ \and Himanshu Tyagi$^4$ \and Pramod Viswanath$^{1,4}$ \and Xuechao Wang$^6$} 

\date{$^1$ Princeton University \\
$^2$ Tsinghua University \\
$^3$ University of Illinois Urbana-Champaign \\
$^4$ Witness Chain \\ $^5$ Eigen Layer \\ $^6$ HKUST \\ ~~\\
\today
}

\begin{document}

\maketitle
\thispagestyle{fancy}
\begin{abstract}
    Large AI models (e.g., Dall-E, GPT4) have electrified the scientific, technological and societal landscape through their superhuman capabilities. These services are offered largely in a traditional web2.0 format (e.g., OpenAI's GPT4 service). As more large AI models proliferate (personalizing and specializing to a variety of domains), there is a tremendous need to have a neutral trust-free platform that  allows the hosting of AI models, clients receiving AI services efficiently, yet in a trust-free, incentive compatible, Byzantine behavior resistant  manner. 
    In this paper we propose  \sakshi, a trust-free decentralized platform specifically suited for AI services. The key design principles of \sakshi are the separation of the data path (where AI query and service is managed) and the control path (where routers and compute and storage hosts are managed)  from the transaction path (where the metering and billing of services are managed over a blockchain). This separation is enabled by a ``proof of inference" layer which provides cryptographic resistance against a variety of misbehaviors, including poor AI service, nonpayment for service, copying of AI models. This is joint work between multiple universities (Princeton University, University of Illinois at Urbana-Champaign, Tsinghua University, HKUST) and two startup companies (Witness Chain and Eigen Layer). 
\end{abstract}
\section{Introduction}

\noindent \textbf{Era of AI}. 
 Artificial Intelligence (AI) has been steadily making progress on a variety of tasks (household tasks by vacuuming robots \cite{roomba, atlas}, playing games -- Chess, Go \cite{silver2017mastering, silver2017masteringgo, silver2018general} -- at superhuman levels, scientific discovery via protein folding predictions \cite{jumper2021highly, evans2021protein}, medical progress by drug discoveries \cite{bostrom2018expanding, strokach2020fast, schneider2020rethinking}), but have broken through the barrier of {\em general intelligence} in recent months with the emergence of a new family of {\em generative} deep learning models --  {\tt GPT4} \cite{openai2023gpt4, bubeck2023sparks} is the prototypical application capturing the world's attention, at a tremendous energy price. {\tt GPT4} has super-human  mastery over natural language, and can comprehend complex ideas, exhibiting proficiency in a myriad of domains such as medicine, law, accounting, computer programming, music, and more. Moreover, {\tt GPT4} is capable of effectively leveraging external tools such as search engines, calculators, and APIs to complete tasks with minimal instructions and no demonstrations, showcasing its remarkable ability to adapt and learn from external resources. Such progress portends AI's forthcoming dominance in mediating (and replacing under several situations) human interactions, and promises AI to be the dominant  energy consuming activity for years to come. \\

\noindent \textbf{Large Generative AI Models}.  An  AI model that is largely representative of the class is {\em  generative AI}, which   creates content that  resembles human-generated ones. These models have attracted considerable interest and popularity due to their impressive capabilities in generating high-quality, realistic images, text, video and music. For instance, large language models (LLMs) like ChatGPT \cite{chatgpt}, Bard \cite{bard}, and LLaMA \cite{touvron2023llama} attain impressive performance on a wide array of tasks and are being integrated in products such as search engines \cite{bing}, coding assistants \cite{copilot} and productivity tools in Google Docs \cite{google_docs}. Further, text-to-image models like StableDiffusion \cite{rombach2022high}, MidJourney \cite{midjoourney}, Flamingo \cite{alayrac2022flamingo}, text-to-music models like MusicLM, \cite{agostinelli2023musiclm} and text-to-video models like Make-a-Video \cite{singer2022make} have shown the immense potential of large multimodal generative AI models.  As large generative AI models continue to evolve, we will witness the emergence of numerous fine-tuned and instruction-tuned models catering to specific use cases (e.g.,  healthcare, finance, law). 
Whilst models grow rapidly, Amazon and Nvidia report that AI inference tasks particularly account for up to 90\% of the computational resource in AI systems, which are much more frequently demanded than AI model training tasks \cite{mcdonald2022great}. In this white paper, we mainly focus on the AI inference tasks, but the flexibility of our layer architecture design allows the market for model training as well.
\\

\noindent \textbf{Current model: Centralized inference}. 
The dominant platform  of serving these large models is through public inference APIs \cite{openai, forefront_ai, ai21}, offered via  by the dominant platform companies of today's economy. For example, the OpenAI API allows users to query models like ChatGPT and DALL-E over a web interface. Although this is a relatively user-friendly option, it is susceptible to the deleterious side-effect of centralization: monopolization. Apart from the rent-seeking aspect of the centralized nature of the service offering, privacy implications loom large: the human interactions mediated by generative AI models is vastly more personal and intrusive than a web browsing and search queries.  Addressing the grand challenge of AI computation via the design of decentralized and programmable platforms is the goal of this paper. \\

\noindent \textbf{Proposed model: Decentralized Inference}. 
In this paper, we propose to decentralize AI inference across servers provided by consumer devices at the grid edge. Decentralized inference can reduce communication and energy costs by leveraging local computation capabilities. This is made possible by utilizing energy-efficient devices located at the edge, which could potentially be powered by renewable energy sources. Crucially, the energy overhead of running large data-centers is largely reduced, simultaneously opening an opportunity to democratize AI whilst limiting its ecological footprint.  Such a decentralized platform would also enable the deployment of a library of large customized models in a scalable manner - users can host in-demand customized models on this decentralized cloud, and earn appropriate rewards.

\begin{figure}
    \centering
    \includegraphics[width=0.8\linewidth]{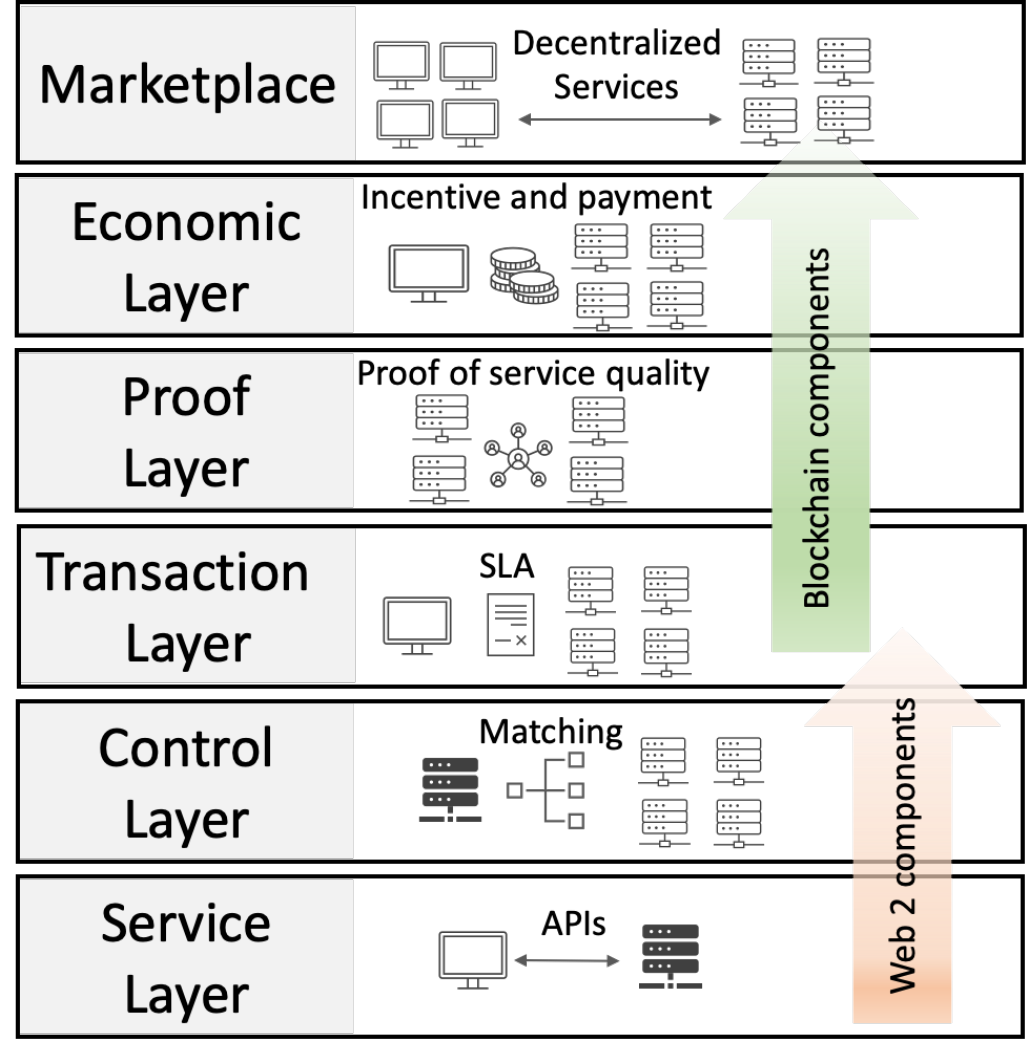}
    \caption{The six layer architecture for Web3.0 services}
    \label{fig:LayerStack}
\end{figure}

Our decentralized AI platform, \sakshi, is populated by a host of different agents: AI service providers, AI clients, storage and compute hosting nodes. A carefully designed incentive fabric stitches the different agents together into an efficient, trustworthy, and economically fruitful AI platform.  Our design of \sakshi is best visualized in terms of a layered architecture (analogous to network stacks).  The layers are enumerated below and visualized in Figure \ref{fig:LayerStack}. 

\begin{enumerate}
    \item {\bf Service layer.} This is the path where the query and response (AI inference) are managed. The goal is to have high throughput and low latency – the goal is to enable user journey similar to a standard web2-like service, with the underlying resources (storage, computation) and economic transaction managed in a decentralized and trustless manner. 
    \item {\bf Control layer.} This is the path where networking and compute/storage load balancing actions are managed. The decentralized AI models are hosted at multiple locations connected via a (potentially peer to peer) network, and our decentralized design borrows from classical web2 content delivery network designs (e.g., Akamai)  while managing the economic transaction also in a decentralized and trustless manner. 
    \item {\bf Transaction layer.} This is the path where billing and metering are conducted. The key is to have this outside the data path and visible to a broader audience (e.g., via commitments on blockchains). Importantly this is trust free crucially enabled via Witness Chain’s transaction layer service (originally designed for decentralized 5G wireless networks \cite{anand2022trust}, but now naturally repurposed for decentralized AI services). 
    \item {\bf Proof layer.} Any dispute in terms of metering and billing are handled here. These proofs also provide resistance to unauthorized usage (e.g., just copying) of AI models. This is definitely outside the data path, but also outside the transaction path. This layer allows the formulation of novel research questions (at the intersection of large AI models, cryptography and security). We highlight three such key questions:  (i) Proof of Inference – where the proof of computation of a specific (deep learning) AI model can be verified;  (ii)  Proof of ownership, fine-tuning and watermarking – where the proof of downstream modification to an AI model can be verified; (iii) Proof of service delivery – where the proof of the delivery of an AI service can be verified at customizable granularities. These dispute resolutions naturally feed into a reputation system (leading to positive incentives for salutary behavior) or crypto economic security via slashing (negative incentives; see next layer). This new research, outlined in detail in this paper, is joint work between multiple universities (Princeton University, University of Illinois at Urbana-Champaign, Tsinghua University, HKUST), and two blockchain startups Witness Chain and Eigen Layer. 
    \item {\bf Economic layer.} So far, the transactions can be handled purely via fiat without the need for a token. This layer explores the benefits of having a token to incentivize participants, both in the transient and long term stages and the corresponding economic benefits therein. → Eigenlayer integration and ideas. 
    \item {\bf Marketplace.} Compositional AI services, in a single atomic transaction, are naturally enabled. The common data shared on the blockchain leads to the creation of a decentralized marketplace for AI services. The supply and demand allows the efficient discovery of prices. Optional in the first version.

\end{enumerate}
\section{Architecture of Decentralized AI Service}

\subsection{Requirements}

We now describe a specific architecture based on the general six layer architecture outlined in the last section, allowing \sakshi to be concrete. Our decentralized AI service is designed to enable an open marketplace for AI models where any user can access inference service offered by multiple, untrusted AI service suppliers. Our goal is to ensure that the user is guaranteed a good quality of service and the suppliers get a fair payment for their service. 

There are several challenges that can hinder bootstrapping and growth of such a decentralized service:
\begin{enumerate}
    \item Individual suppliers may not be able to attract enough clients;
    \item The supplier may not apply a good model and return low quality results;
    \item The client may not pay after getting the service.
\end{enumerate}

Each of these challenges is addressed by our decentralized AI service model:
\begin{enumerate}
    \item We allow an aggregator to collectively offer service on behalf of multiple suppliers. The aggregator and suppliers engage in an SLA implemented as a smart contract to ensure that each gets a fair share of the revenue.
    \item We have a proof system for quality of AI services to ensure that suppliers provide the promised quality of service. The proof is implemented through a challenge-response setup executed using a decentralized pool of challenger nodes.
    \item We have smart contracts and payment channels to implement scalable and reliable payment service for the suppliers. This will be supported by an objective dispute resolution mechanism to ensure that suppliers can get paid if they deliver service.
\end{enumerate}

\subsection{The six layer architecture with Witness Chain }

\begin{figure}
    \centering
    \includegraphics[width=\linewidth]{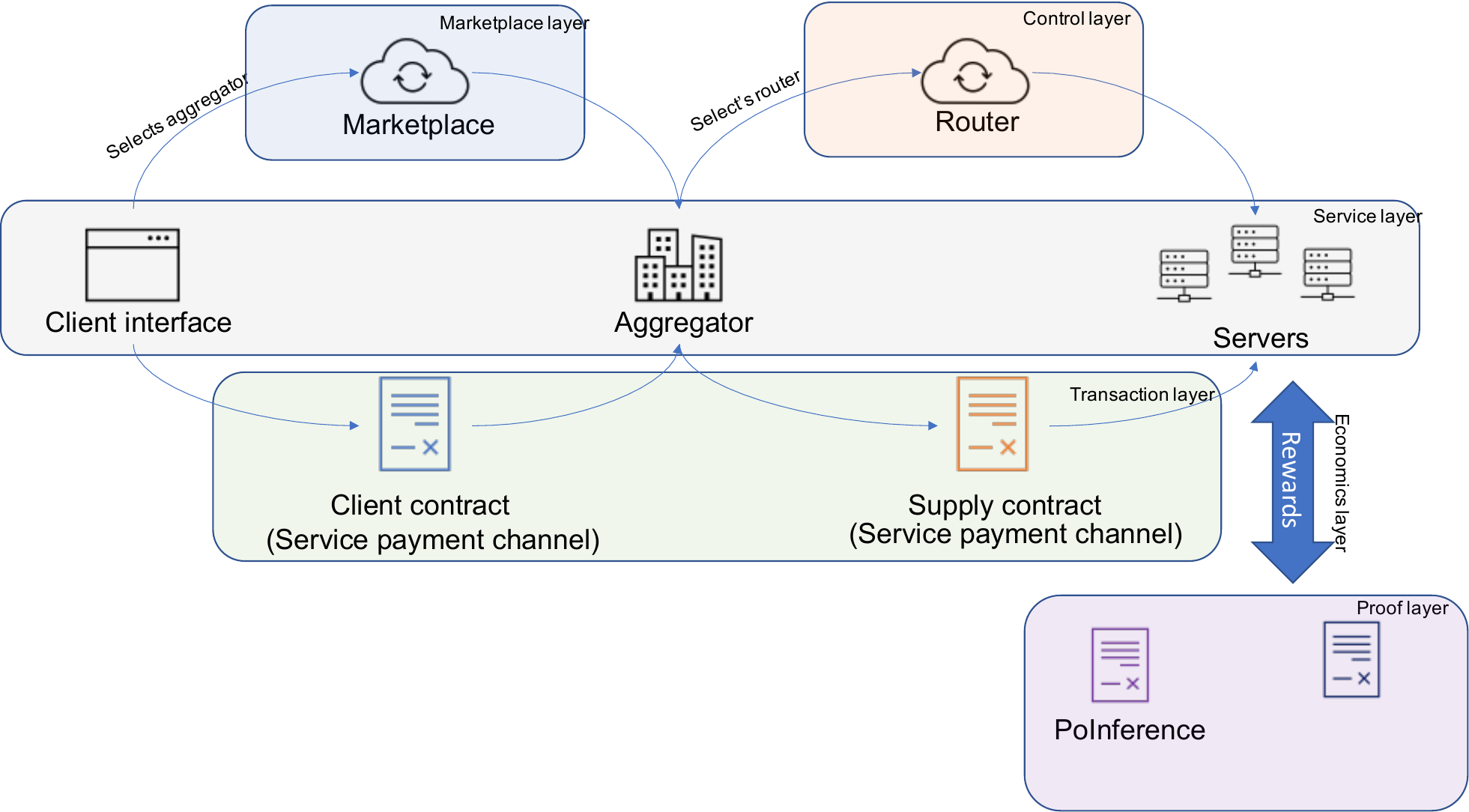}
    \caption{\sakshi - Decentralized AI service architecture}
    \label{fig:LayerArchitecture}
\end{figure}

These functionalities of \sakshi are enabled using the  architecture in Figure \ref{fig:LayerArchitecture}.  

At the top is the marketplace, a decentralized two-sided platform for buying and selling AI services. A client (user) comes to our marketplace and places an order to access inference service from an aggregator. Both agree on an SLA which contains terms for quality of service and payments. 

Next comes the service layer that provides the APIs for clients to make inference requests to the aggregators. This request is appropriately passed to a matching supplier server using a router deployed as a part of the control layer. Both service and control layer are reminiscent of standard web 2.0 services with multiple servers, with the caveat that the supplier servers can now be hosted by different entities with their own business incentives and without any pre-existing reputation. These servers are bound to an SLA between them and the aggregator. 

All the SLAs that govern the service-payment rules between different parties are deployed as smart contracts as a part of the transaction layer, a decentralization middleware provided by Witness Chain \cite{witnesschain}. The Witness Chain transaction layer not only hosts and provides interfaces for the SLA smart contracts, but also provides state channels to maintain the payment and service state for interacting client, aggregator and supplier. Furthermore, it provides a dispute resolution framework to ensure that the client completes the payment after availing the service. 

 Finally, a proof layer deploys an appropriate Proof of Inference to ensure that the suppliers are using models agreed upon in the SLA. This challenge and verification for this proof is executed by a pool of challengers, Witnesses, provided by Witness Chain. These proofs interact with the transaction layer to ensure service quality promised in the SLA. The Witness Chain challenger nodes executing these proofs are incentivised by Witness Chain using a part of service payment. Witness Chain, in turn, provides a programmable layer for choosing the challenger nodes which can be used to specify how decentralized the challenger pool should be and how well-provisioned each challenger node needs to be. 

A detailed description of each layer is provided in Section \ref{sec:detailed}; the interactions discussed above are depicted at a high level in Figure \ref{fig:ServiceSteps} below. 

\begin{figure}
    \centering
    \includegraphics[width=\linewidth]{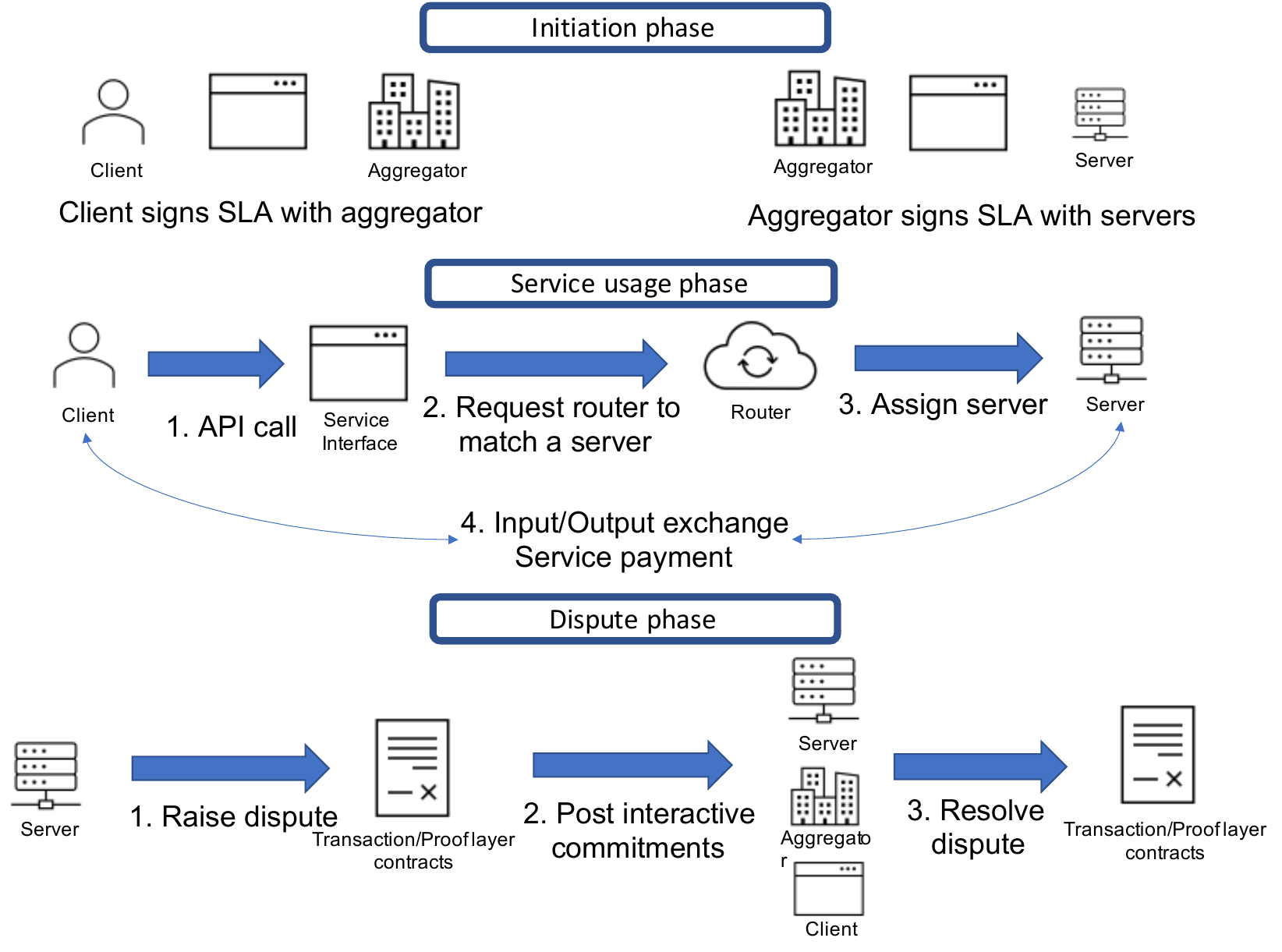}
    \caption{Various steps in using \sakshi}
    \label{fig:ServiceSteps}
\end{figure}

\subsection{The economic layer with Eigen Layer}

All entities in the above ecosystem are incentivized to do their job fairly because of the economics underlying the SLA and the incentive system for the challengers. Often, each new blockchain ecosystem launches its own token to provide this cryptoeconomic security. However, this new token may not gain the necessary volume and spread to enforce reasonable security in the early stages, resulting in failure to bootstrap for the ecosystem. 

This problem was solved recently by Eigen Layer \cite{eigenlayer} which provides a framework for using Ethereum cryptoeconomic security by engaging Ethereum validators. Witness Chain integrates with Eigen Layer and uses Eigen Layer operators as challengers to extend Ethereum security to the decentralized AI marketplace. The challengers running the Proof of Inference, the ultimate root of trust in service quality, would have staked/restaked Eth using Eigen Layer.  Witness Chain deploys an additional proof of custody \cite{witnesschain} to ensure that these challengers are being diligent in their job, lest their stake  be slashed. Putting the restaking framework of Eigen Layer together with the proof of diligence/custody by Witness Chain provides a comprehensive economic security layer for \sakshi.

\section{Detailed Description of Each Layer} \label{sec:detailed}

\subsection{Service layer}

\begin{figure}
    \centering
    \includegraphics[width=0.9\linewidth]{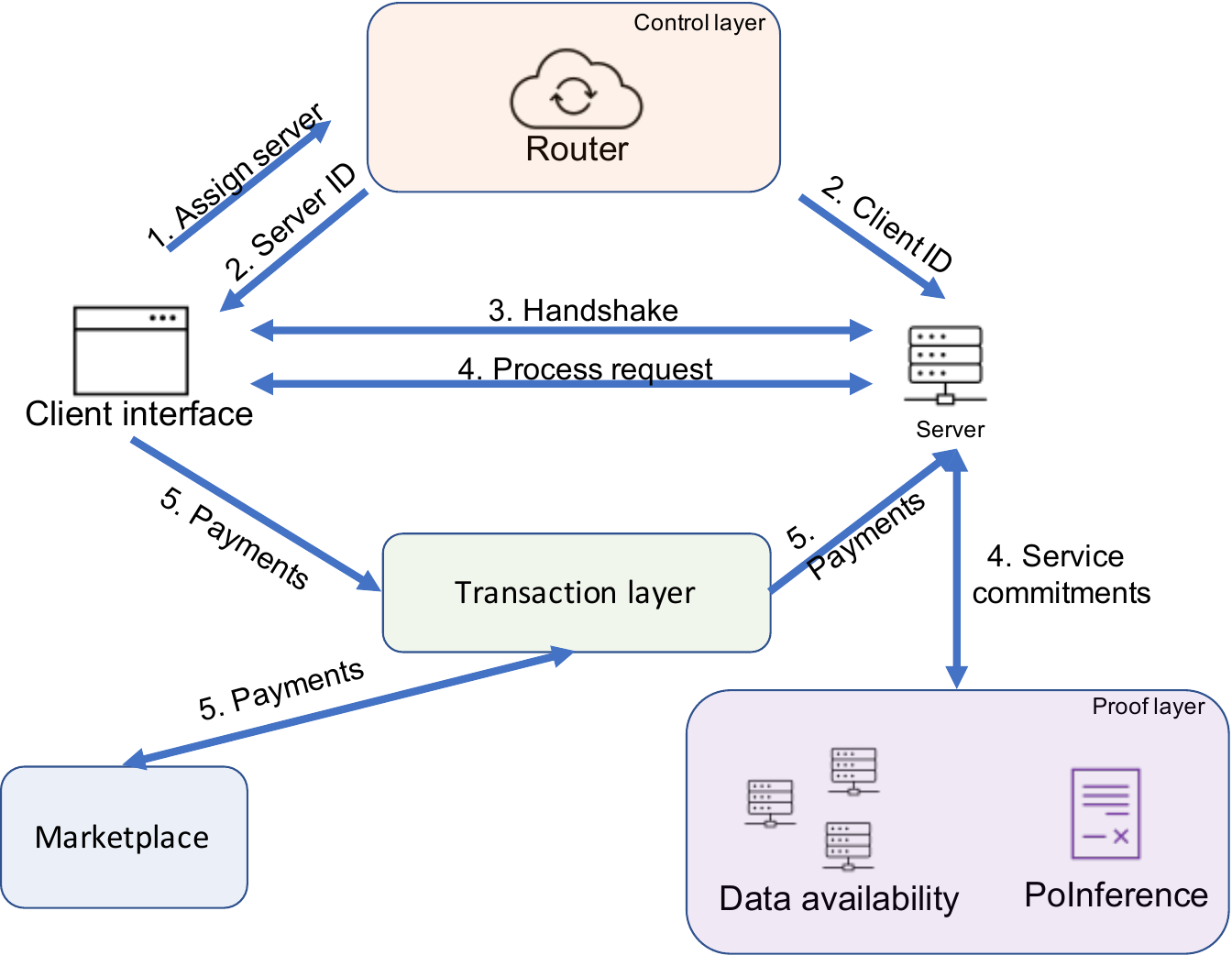}
    \caption{Service Layer overview}
    \label{fig:ServiceLayer}
\end{figure}

The service layer enables the infrastructure for ML inference queries and is responsible for committing service information to the proof layer. This layer is equivalent to a Web2 server-client architecture with some modifications to support the proof framework. An instantiation of this layer creates a connection between a client and a server to exchange data and makes the server’s compute available through agreed-upon Inference APIs. The service layer works in conjunction with other layers in the infrastructure as depicted in Figure \ref{fig:ServiceLayer} 
and described below:\\

\noindent {\bf Server Assignment:} The client requests the control layer to assign a server for an AI model, and the control layer notifies the client of the server’s ID and address. It also notifies the server of an incoming connection from the client. \\

\noindent {\bf Service exchange:}  The client establishes a connection with the server using the address provided by the control layer. Both server and client verify through the transaction layer if an SLA path exists between them through the common aggregator; if such a path exists, both parties implicitly agree on the trade. The client sends inference requests using the server’s API endpoint; the client signs the request for use in dispute resolution if the need arises. The server processes the requests and sends the output data back to the client as the response; the server might submit a commitment to the delivered response on a DA layer at a later stage if the need arises for dispute resolution. Per service of a single unit of inference - a single API request, the server anticipates a micropayment as dictated by its SLA. A request is made to the transaction layer, which then sends payments from the client to the aggregator and from the aggregator to the server. The server proceeds to serve the subsequent request from the client only if the payment for the previous request is processed.\\

\noindent {\bf Service dispute witnesses:} The data exchanged in the service layer is used as a witness in case a payment dispute arises, such as a client not paying for the AI inference service delivered. The signed inference requests, output data committed to a DA layer, and the previous exchanged micropayment will be used for dispute resolution, as discussed in detail in the following sections on the Transaction and Proof layers. 

\subsection{Control Layer}

\begin{figure}
    \centering
    \includegraphics[width = 0.9\linewidth]{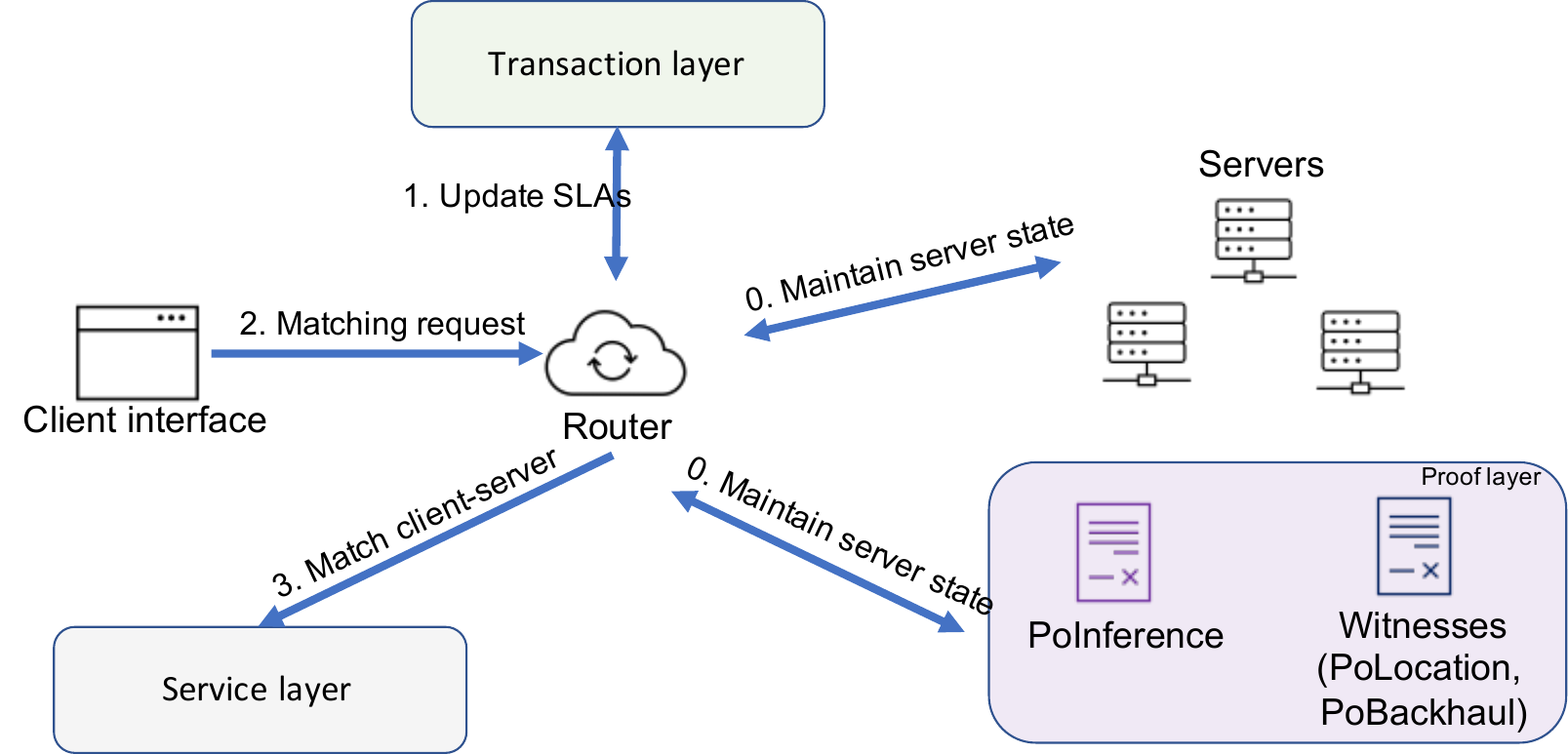}
    \caption{Control layer overview}
    \label{fig:controlLayer}
\end{figure}

The control layer is responsible for matching clients and servers. This layer consists of a set of routers that maintains the state of all servers subscribed to it. It performs load balancing by allocating client requests to servers that optimize cost measured in latency, compute cost, and compliance to SLAs. Servers can subscribe to a router of their choice, and clients can select a router of their choice. The control layer works in conjunction with other layers as depicted in figure \ref{fig:controlLayer} and described below:\\

\noindent {\bf Server state maintenance:} Router maintains a server network state consisting of the following non-exhaustive set of variables:
\begin{itemize}
    \item Server model capacity: The set of AI models that the server can compute inference on 
    \item Server hardware capacity: The compute capacity of each server 
    \item Server request load: The number of clients the server is currently connected to at the service layer 
    \item Server location: Verified server location from the proof layer
\end{itemize}
Some of these variables require the routing trusting server’s claims - these will be used for soft constraints in routing, whereas other variables such as location will be verified through the proof layer - these can be used for hard constraints such as geo-restricting the inference compute. \\

\noindent {\bf SLA state maintenance:}  The router maintains the state of SLAs signed at the transaction layer between client-aggregators and aggregator-servers so that it can match clients to servers that share a common aggregator. The router watches the transaction layer contracts for events to register or de-register SLAs. \\

\noindent {\bf Client-server matching:} The client submits a request specifying the type of server it would like to be matched to - this request consists of parameters such as model id, location boundary, server uptime, etc. The router runs a matching logic to select a server best suited for that model at that time by utilizing the server state and the SLA state. The router then notifies the service layer to establish a connection between the client and the servers and the transaction layer to anticipate payments through their common aggregator. 

Note on fairness: A malicious router can unfairly route requests leading to a loss in revenue for some servers; if a server sees such behavior, it will migrate to another router that provides better revenue by providing fair routing. This market dynamic facilitates fairness in routing.

\subsection{Transaction Layer}

The transaction layer is responsible for payment to servers and intermediaries for delivering their service. 

\subsubsection{Necessity of an integrated transaction layer}

Decentralized platforms generate supply by incentivizing and compensating an extensive network of parties - termed suppliers. The platform can be considered a marketplace for the service supply chain, with service flowing from suppliers (servers) to intermediaries and finally to consumers and compensation flowing the other way. A compensation system is, therefore, a critical part of a decentralized service-oriented platform. 

Compensation for providing services is already an integral part of existing centralized platforms such as Uber, AirBnB, and Amazon; however, the billing systems used for their decentralized counterpart need to be composable with the trustless and programmable service framework that decentralized platforms enable. Decentralized platforms need the billing system to support automated smart contract-initiated dispute resolution and high-speed dispersion of funds, as we will see next. The transaction layer incorporates the web3 equivalent of a billing system. The transaction layer ties the billing of a service with a Service Level Agreement (SLA) that codifies the terms of service and payment, and ensures that metering for the SLA is consistent with the service delivered.

\subsubsection{Scalability solutions}
Decentralized AI platforms cannot rely on the assumption of trust between a server and a client since either party may be too small to be bound by the principles of reputation maintenance or legal agreements. Thus, they need to be constantly in consensus about the amount of inference service delivered and payment for such service. A requirement for achieving this consensus is that it must be achieved per delivery of an inference service unit - a query. All parties involved in service delivery must agree on the service delivered and settle payment for that service delivered at frequent intervals. This requirement necessitates a high throughput, low latency payment system. 

Consensus literature is rich in solutions to scale payment p ranging from sharding, rollups, and sidechains to payment channels. Our payment system should ideally satisfy the following properties:
\begin{itemize}
    \item High throughput of payments 
    \item Low latency between payment initiation and confirmation 
    \item Scale throughput with the number of supply or demand side participants 
    \item Payment per service delivery is not public information and may only be shared between the supplier, consumer, and the chosen intermediaries. 
\end{itemize}

State channels and payment channels satisfy all the above requirements. 
Modeling a decentralized AI platform, we observe that a single client will interact with multiple servers to query for different models and use different suppliers for inter-session privacy. The requirement for managing a state channel across multiple servers is not scalable. Hence we choose a payment channel approach to build the transaction layer’s payment system. We will have a payment channel between a client and an aggregator intermediary and another between the aggregator intermediary and server, enabled by SLA chaining. Figure \ref{fig:transactionLayer} depicts the interaction of transaction layer components with other layers, with details on the architecture below:

\begin{figure}
    \centering
    \includegraphics[width = 0.9\linewidth]{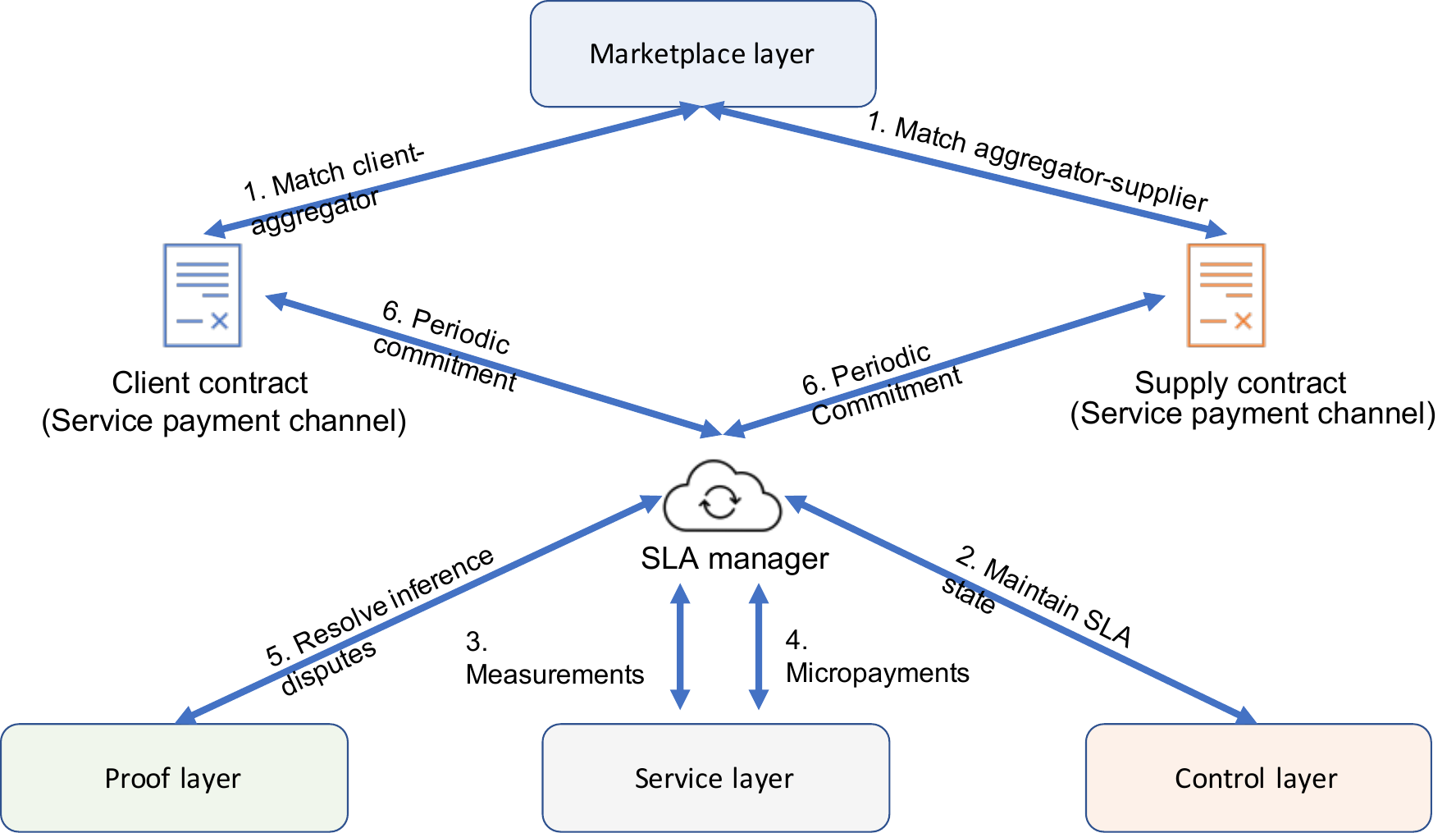}
    \caption{Transaction layer overview}
    \label{fig:transactionLayer}
\end{figure}

\subsubsection{Architecture overview}

The transaction layer encompasses SLAs that any two parties agree on, an SLA manager that converts service measurements to payments using SLA, SLA clients running on machines of both parties fetching data from the measurement gateway, and a blockchain wrapper for posting transactions. These components are described in detail below:

\noindent {\bf Service contracts:} Service contracts consist of two components: A SLA that both the transacting parties agree on and a unidirectional payment channel with funds flowing from the service consumer to the supplier. For the AI platform there exists two consumer - supplier pairs: (i) Client - Aggregator and (ii) Aggregator - Server. The SLA is codified based on a SLA4OpenAPI standard \cite{sla4openAPI} and maps service usage to a payment. SLAs for AI application maps (model type, input size, output size) to token payment amount. The unidirectional payment channel is set up with an escrow from the consuming party to supplying party and set’s terms of delegation of payment keys to an intermediary SLA manager. 
\looseness=-1 

\noindent {\bf SLA manager:} SLA manager
end clients are given to run a codebase that signs micropayments or delegate it to an application running on the cloud: SLA manager. SLA manager receives signed measurements from the consumer and supplier’s SLA client and converts that to an appropriate payment amount by signing a micropayment and sending funds on the payment channel on behalf of the consumer.\\

\noindent {\bf SLA client and measurement gateway:} SLA client and measurement gateway are components that run on the end devices of the consumer and supplier. The measurement gateway interprets the service messages and converts them into service units. For AI applications, these would be the model requested, input size, and output size. The SLA client fetches this information from the measurement gateway, signs it with the key codified in the service contract, and sends it to the SLA manager; optionally, the SLA client (on the consumer end) can convert the measurement to a micropayment themselves and forward it to the supplier. \\
\looseness=-1

\noindent {\bf Blockchain wrapper} This component runs on the SLA manager and SLA client. It is responsible for broadcasting and listening to on-chain transactions such as payment channel start, termination, and dispute messages on-chain. The blockchain wrapper is compatible with multiple blockchains such as Ethereum, Polygon, Solana, and all EVM-compatible rollups. \\

\subsubsection{Dispute-compatibility}

\begin{figure}
    \centering
    \includegraphics[width = \linewidth]{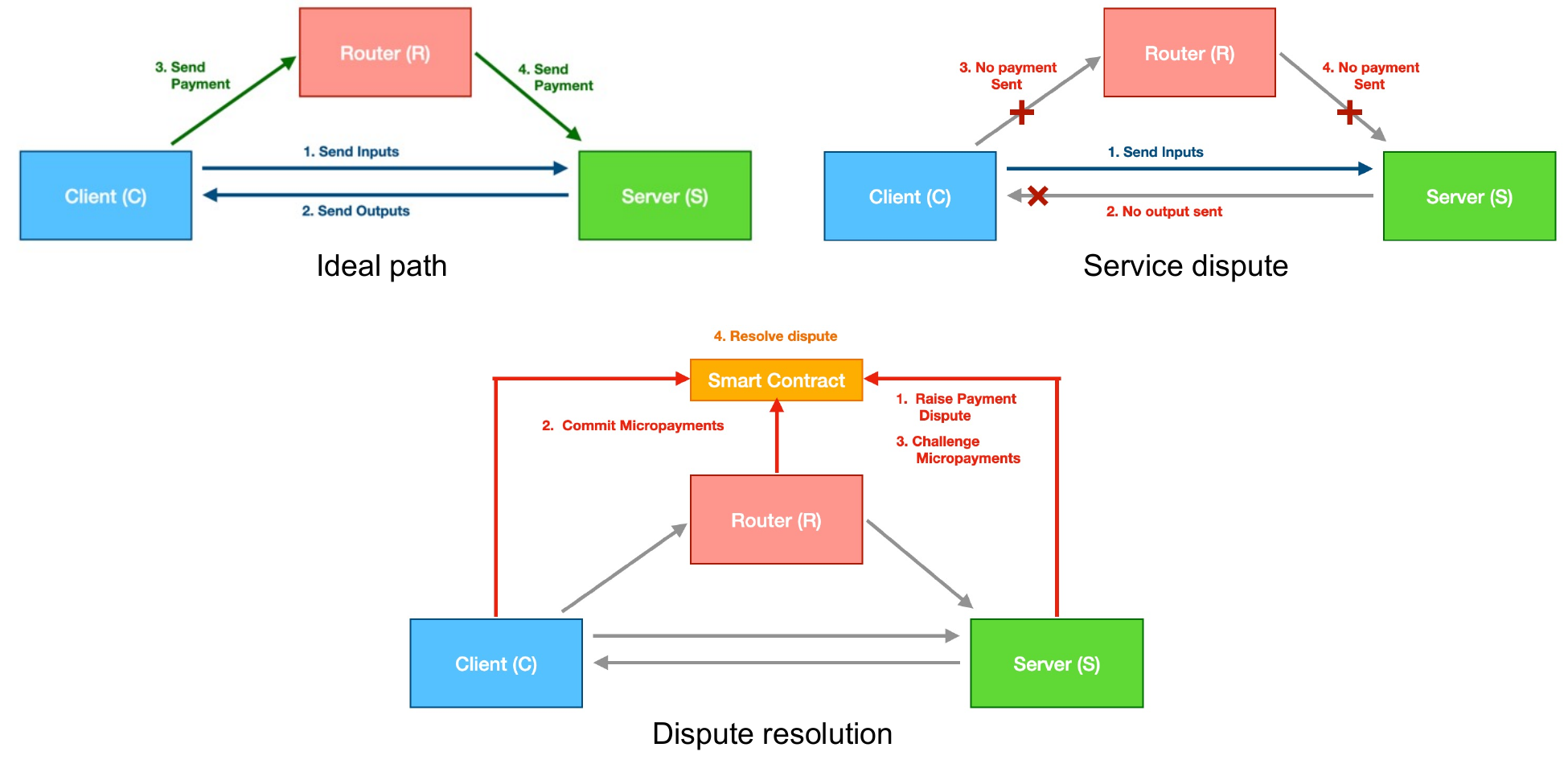}
    \caption{Utilizing transaction layer payments for service dispute resolution}
    \label{fig:TLDisputeResolution}
\end{figure}

\sakshi utilizes a post-service payment model - Payment disputes can emerge when a supplier claims non-receipt of payment for a service unit (a single AI inference). The associated micropayment can serve as a proof of payment to resolve such disputes. Micropayments in unidirectional payment channels typically consist of a signed commitment of the total payable amount. To render these payment channels to be dispute-compatible, we need to augment them with additional parameters. Firstly, the micropayment should include a unique ‘requestID’ that corresponds to the disputed inference. Secondly, it should contain the hash of the preceding micropayment, which can be validated using a nonce - a counter incremented with each successive micropayment. 
To resolve a payment dispute raised by the server, the payer can commit the associated micropayment. Additionally, the preceding micropayment must also be committed, to calculate the amount payable for the disputed service unit. Depending on who is deemed to be correct, the dispute can be settled on-chain from the existing balance in the payment channel. Our dispute resolution protocol also addresses other scenarios, such as disputes raised by a malicious server without providing service, and inconsistent micropayment commitments. Figure \ref{fig:TLDisputeResolution} depicts an example flow of utilizing payment channel commitments for service dispute resolution.

\subsection{Proof Layer}

The proof layer, operating outside the data and transaction paths, provides a way to resolve various disputes in \sakshi, utilizing blockchains as a immutable and trusted medium to read and write service states. A variety of disputes can arise in the AI service and ``proof" systems to provide cryptographic resolution mechanisms address the corresponding issues. In this paper, we focus on two  categories of proofs, each responding to different types of disputes. 
\begin{itemize}
    \item Proof of Inference, a proof of correct computation on a prescribed (and open) AI model, mediates disputes of correct inference;
    \item Proof of Model-ownership, a proof of how closely two AI models are related to each other and whether one AI model is a clone or a fine-tuned version of the other, mediates potential disputes related to intellectual property held by the owner of an AI model. 
\end{itemize}

\begin{figure}
    \centering
    \includegraphics[width = \linewidth]{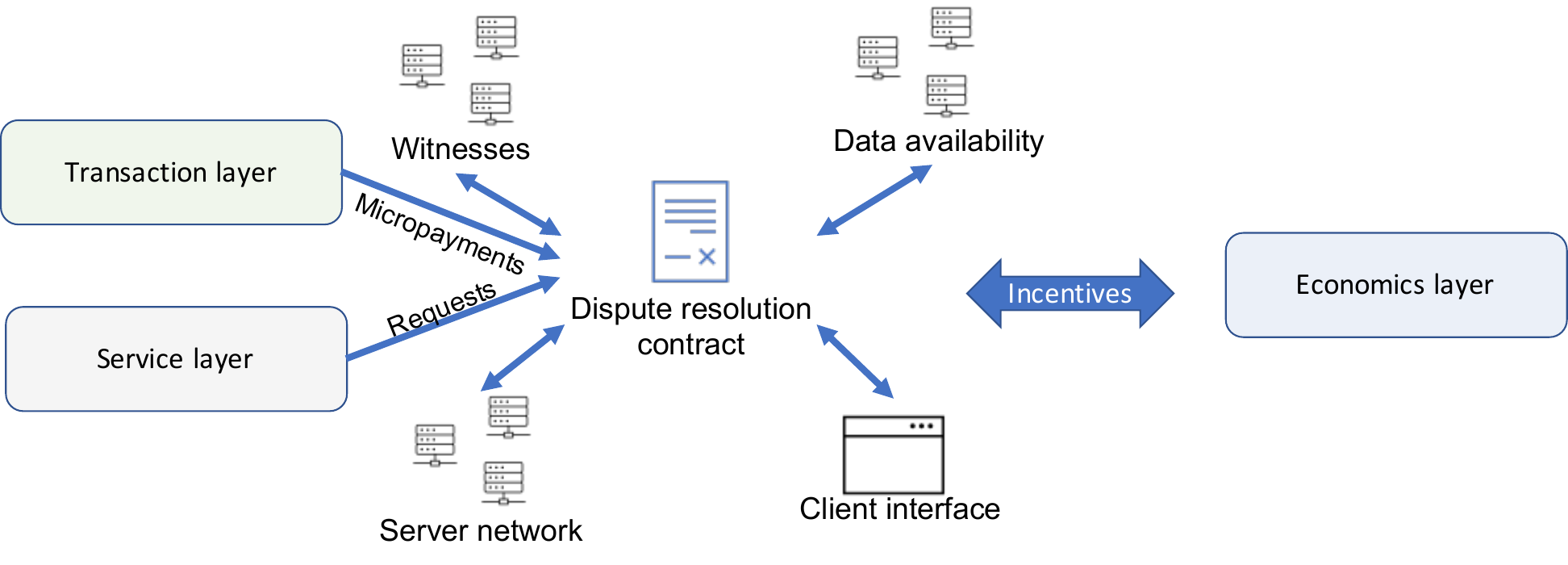}
    \caption{Proof layer overview}
    \label{fig:proofLayer}
\end{figure}

Figure \ref{fig:proofLayer} depicts the interaction of the dispute resolution contract in the proof layer with the rest of the platform layers. A detailed description of the individual proof follows. 

\subsubsection{Proof of Inference}
A crucial aspect of decentralized inference platforms is the presence of incentives that encourage honest participation in the protocol while discouraging malicious actors. An essential component of this incentive design is addressing the problem of provably verifying computations executed by untrusted servers. Various design choices are available to enable such proof of inference, with several emerging research directions.

One such line of research involves the application of zero-knowledge proofs (ZKP) to verify AI model execution \cite{kang2022scaling}. However, this approach is extremely computationally intensive, necessitating concessions such as quantization, which leads to lower accuracy. Furthermore, generating ZKPs for modern, large-scale generative AI models is currently impractical. 

An alternative strategy is to adopt an optimistic approach. In this scheme, the server commits the hash of the generated output, and the system assumes the off-chain inference to be accurate. If a participant (``challenger") doubts the inference's correctness, they can contest its validity by submitting a fraud proof. This proof can be generated using a verification oracle that can re-run the model and determine the accuracy of the server's or challenger's claim. However, since these oracle nodes may have limited computational capabilities, recomputing the entire neural network forward pass is prohibitively expensive and inefficient.

To address this issue, we propose a method inspired by the bisection scheme employed in the optimistic rollup Arbitrum~\cite{kalodner2018arbitrum}. A key observation is that AI models can be viewed as a sequence of functions, such as layers in a neural network.
$$  f(x) = y \quad \rightarrow \quad f_n(f_{n-1}(f_{n-2}(...f_2(f_1(x))...))) = y$$
When there is a discrepancy between the outputs of a server and a challenger, we can employ an interactive bisection scheme to identify a single function—the first layer in the AI model where the outputs of the two parties differ. By implementing this system, oracle nodes only need to compute and verify a single layer of the network, significantly reducing costs and making the verification of extremely large models feasible. Indeed, deterministic AI inference is a prerequisite for such schemes, which is attainable by fixing the random state.

\begin{figure}
    \centering
    \includegraphics[width=\linewidth]{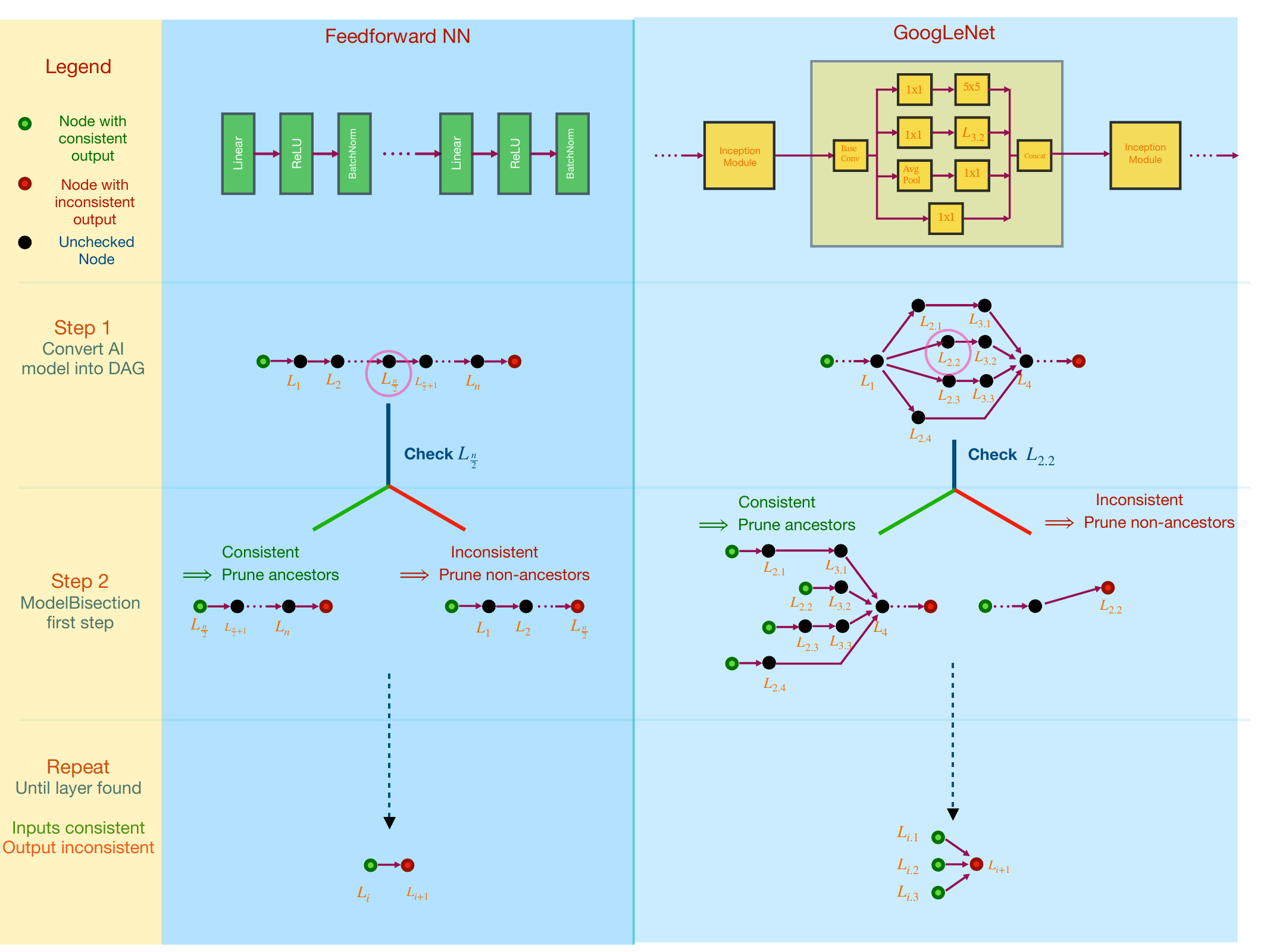}
    \caption{Model bisection}
    \label{fig:modelbisection}
\end{figure}

We illustrate our ModelBisection algorithm in Figure \ref{fig:modelbisection}, that identifies the earliest layer of the AI model where the inputs align for both parties, but the resulting outputs diverge, while minimizing the number of interactive steps involved. In case of a sequential model (left), one can use a form of binary search - if the output of a queried layer (typically the midpoint) is inconsistent between the parties, we recursively bisect the first half of the node sequence. Otherwise, we eliminate the first half, and recursively bisect the second half of the sequence. Each bisection step eliminates half of the remaining candidates for the faulty layer. After a logarithmic number of iterations, we locate a layer whose input is consistent, yet the parties produce differing outputs. 

However, the computations within an AI model are not simply sequential but rather form a Directed Acyclic Graph (DAG) structure. Consequently, the bisection mechanism used for sequential networks cannot be directly applied to AI models. We demonstrate our approach, \emph{ModelBisection}, on an Inception block of GoogLeNet \cite{szegedy2015going} as depicted in Figure \ref{fig:modelbisection} (right). Suppose we select the node $n_1 = L_{2.2}$ in the DAG for output verification. Both parties compute and share the intermediate output of layer $L_{2.2}$. If the outputs are equal, we prune all ancestor nodes of this node in the DAG from consideration (as their outputs would have to be consistent). If, however, the outputs differ, we eliminate all non-ancestor nodes of this node in the DAG (since one of outputs among ancestors must be inconsistent). We keep track of the identified consistent and inconsistent nodes, and continue this process until we reach a single layer where the inputs are consistent between the parties, but the outputs differ. We employ a greedy strategy to select the node in the digraph such that it is split in the most balanced way. We choose the node which maximizes $\min\{|x|, n - |x|\}$, where $|x|$ is the number of ancestors of node $x$, and n is the total number of nodes in the current digraph. This score can be interpreted as the least number of nodes that would be eliminated as potential candidates for the first point of divergence, when $x$ is queried, thus minimizing the number of ModelBisection rounds. It's noteworthy that even in large foundation models, the ModelBisection approach can pinpoint a single layer of divergence in a very small number of iterations. For example, in the case of the 13 billion parameter LLaMA model \cite{touvron2023llama},  fewer than ten iterations suffice. Finally we observe that the bisection subroutine bears similarity to the one utilized by GitHub in \emph{git bisect}, which aids in identifying the first faulty entry in the DAG of commits and merges.

\subsubsection{Proof of Model ownership}

A decentralized AI marketplace comprises three main entities - model owners who collect datasets and train or finetune AI models, compute-rich servers, and end-users. As opposed to current open-source model hosting solutions, decentralized marketplaces can allow incentivizing model creators by rewarding them a percentage of the inference fee when their models are utilized.
However, such an incentive design is susceptible to model copying attacks, where a malicious actor can copy, slightly modify, and profit from the hosted models at the cost of the model creators. Therefore, a robust mechanism for model ownership resolution becomes a crucial prerequisite for decentralized AI marketplaces.

One promising solution for a proof of model ownership is by embedding a watermark in the neural networks during the training phase. To be effective, a DNN watermarking scheme must fulfill several criteria: it should be functionality-preserving, meaning the watermark embedding must not impact model performance. The watermark must be robust, and be extractable from any transformed model (e.g., through weight scaling or finetuning). Additionally, a watermarked model should remain indistinguishable from a non-watermarked model to potential adversaries. Moreover, a watermark must be resistant to ambiguity attacks - false claims of existence of a different watermark.

Various watermarking schemes have been proposed in research literature. Parameter encoding methods \cite{uchida2017embedding, darvish2019deepsigns, fan2019rethinking}, integrate a watermark directly into the model's parameters. For classification models, an alternate method involves backdooring, which involves assigning incorrect labels to examples in a trigger set, and this can be used as a watermark \cite{adi2018turning, szyller2021dawn}. Additionally, task-specific and model-specific watermarking methods have been proposed \cite{fernandez2023stable, zhao2023recipe, christ2023undetectable, kirchenbauer2023watermark}. Nonetheless, the robustness of existing methods against model copying has been questioned by recent attacks \cite{lukas2022sok, yan2023rethinking, liu2023false}, highlighting an unresolved research challenge.

Notably, in most watermark extraction algorithms, information about the watermark location or the trigger examples are revealed during the verification process. This knowledge facilitates easier watermark removal and ambiguity attacks. Therefore, in our system a trusted judge is required to resolve model ownership disputes. Model creators must embed watermarks in their models, and commit a commitment of the watermark on the blockchain. The judge must be able to verify the existence of watermarks using the extraction algorithm, which may be task and model-specific. Such a proof of model ownership can ensure the non-feasibility of profiting from stolen models within the decentralized marketplace. However, it does not prevent an adversary from copying a model and using it outside this system (eg - via a black-box api). Such acts can be deterred by licensing the model’s use only in this marketplace, and resorting to legal means if necessary.

\subsection{Summary} Proofs of inference and ownership are two examples of a broader family of protocols providing Byzantine resistance in \sakshi. Even here, we have worked more to describe the problems rather than the solutions -- a call to arms from the scientific community. As the platform evolves and participation rises, the attack space could also expand opening the door for new and different kinds of proof systems (e.g., proof of custody; proof of infrastructure hosting the AI models). 

\bibliography{ref}

\begin{thebibliography}{10}

\bibitem{roomba}
iRobot.
\newblock Roomba robot vacuums.
\newblock \url{https://www.irobot.com/en\_US/roomba.html}.
\newblock Accessed: 2023-03-23.

\bibitem{atlas}
Boston Dynamics.
\newblock The most dynamic humanoid robot.
\newblock \url{https://www.bostondynamics.com/atlas}.
\newblock Accessed: 2023-02-01.

\bibitem{silver2017mastering}
David Silver, Thomas Hubert, Julian Schrittwieser, Ioannis Antonoglou, Matthew
  Lai, Arthur Guez, Marc Lanctot, Laurent Sifre, Dharshan Kumaran, Thore
  Graepel, et~al.
\newblock Mastering chess and shogi by self-play with a general reinforcement
  learning algorithm.
\newblock {\em arXiv preprint arXiv:1712.01815}, 2017.

\bibitem{silver2017masteringgo}
David Silver, Julian Schrittwieser, Karen Simonyan, Ioannis Antonoglou, Aja
  Huang, Arthur Guez, Thomas Hubert, Lucas Baker, Matthew Lai, Adrian Bolton,
  et~al.
\newblock Mastering the game of go without human knowledge.
\newblock {\em nature}, 550(7676):354--359, 2017.

\bibitem{silver2018general}
David Silver, Thomas Hubert, Julian Schrittwieser, Ioannis Antonoglou, Matthew
  Lai, Arthur Guez, Marc Lanctot, Laurent Sifre, Dharshan Kumaran, Thore
  Graepel, et~al.
\newblock A general reinforcement learning algorithm that masters chess, shogi,
  and go through self-play.
\newblock {\em Science}, 362(6419):1140--1144, 2018.

\bibitem{jumper2021highly}
John Jumper, Richard Evans, Alexander Pritzel, Tim Green, Michael Figurnov,
  Olaf Ronneberger, Kathryn Tunyasuvunakool, Russ Bates, Augustin
  {\v{Z}}{\'\i}dek, Anna Potapenko, et~al.
\newblock Highly accurate protein structure prediction with alphafold.
\newblock {\em Nature}, 596(7873):583--589, 2021.

\bibitem{evans2021protein}
Richard Evans, Michael O’Neill, Alexander Pritzel, Natasha Antropova, Andrew
  Senior, Tim Green, Augustin {\v{Z}}{\'\i}dek, Russ Bates, Sam Blackwell,
  Jason Yim, et~al.
\newblock Protein complex prediction with alphafold-multimer.
\newblock {\em BioRxiv}, pages 2021--10, 2021.

\bibitem{bostrom2018expanding}
Jonas Bostr{\"o}m, Dean~G Brown, Robert~J Young, and Gy{\"o}rgy~M Keser{\"u}.
\newblock Expanding the medicinal chemistry synthetic toolbox.
\newblock {\em Nature Reviews Drug Discovery}, 17(10):709--727, 2018.

\bibitem{strokach2020fast}
Alexey Strokach, David Becerra, Carles Corbi-Verge, Albert Perez-Riba, and
  Philip~M Kim.
\newblock Fast and flexible protein design using deep graph neural networks.
\newblock {\em Cell systems}, 11(4):402--411, 2020.

\bibitem{schneider2020rethinking}
Petra Schneider, W~Patrick Walters, Alleyn~T Plowright, Norman Sieroka,
  Jennifer Listgarten, Robert~A Goodnow~Jr, Jasmin Fisher, Johanna~M Jansen,
  Jos{\'e}~S Duca, Thomas~S Rush, et~al.
\newblock Rethinking drug design in the artificial intelligence era.
\newblock {\em Nature Reviews Drug Discovery}, 19(5):353--364, 2020.

\bibitem{openai2023gpt4}
OpenAI.
\newblock Gpt-4 technical report, 2023.

\bibitem{bubeck2023sparks}
S{\'e}bastien Bubeck, Varun Chandrasekaran, Ronen Eldan, Johannes Gehrke, Eric
  Horvitz, Ece Kamar, Peter Lee, Yin~Tat Lee, Yuanzhi Li, Scott Lundberg,
  et~al.
\newblock Sparks of artificial general intelligence: Early experiments with
  gpt-4.
\newblock {\em arXiv preprint arXiv:2303.12712}, 2023.

\bibitem{chatgpt}
Introducing chatgpt, 2022.
\newblock Retrieved March 14, 2023, from https://openai.com/blog/chatgpt.

\bibitem{bard}
Google.
\newblock {BARD}.
\newblock
  \url{https://blog.google/technology/ai/bard-google-ai-search-updates/}.

\bibitem{touvron2023llama}
Hugo Touvron, Thibaut Lavril, Gautier Izacard, Xavier Martinet, Marie-Anne
  Lachaux, Timoth{\'e}e Lacroix, Baptiste Rozi{\`e}re, Naman Goyal, Eric
  Hambro, Faisal Azhar, et~al.
\newblock Llama: Open and efficient foundation language models.
\newblock {\em arXiv preprint arXiv:2302.13971}, 2023.

\bibitem{bing}
Yusuf Mehdi.
\newblock Reinventing search with a new ai-powered microsoft bing and edge,
  your copilot for the web.
\newblock
  \url{https://blogs.microsoft.com/blog/2023/02/07/reinventing-search-with-a-new-ai-powered-microsoft-bing-and-edge-your-copilot-for-the-web/}.

\bibitem{copilot}
Github CoPilot.
\newblock Your ai pair programmer is leveling up.
\newblock \url{https://github.com/features/preview/copilot-x}, 2023.
\newblock Accessed: 2023-03-24.

\bibitem{google_docs}
Google Cloud.
\newblock The next generation of ai for developers and google workspace.
\newblock
  \url{https://blog.google/technology/ai/ai-developers-google-cloud-workspace/},
  2023.
\newblock Accessed: 2023-03-24.

\bibitem{rombach2022high}
Robin Rombach, Andreas Blattmann, Dominik Lorenz, Patrick Esser, and Bj{\"o}rn
  Ommer.
\newblock High-resolution image synthesis with latent diffusion models.
\newblock In {\em Proceedings of the IEEE/CVF Conference on Computer Vision and
  Pattern Recognition}, pages 10684--10695, 2022.

\bibitem{midjoourney}
Midjourney.
\newblock \url{https://www.midjourney.com}.
\newblock Accessed: 2023-03-23.

\bibitem{alayrac2022flamingo}
Jean-Baptiste Alayrac, Jeff Donahue, Pauline Luc, Antoine Miech, Iain Barr,
  Yana Hasson, Karel Lenc, Arthur Mensch, Katie Millican, Malcolm Reynolds,
  et~al.
\newblock Flamingo: a visual language model for few-shot learning.
\newblock {\em arXiv preprint arXiv:2204.14198}, 2022.

\bibitem{agostinelli2023musiclm}
Andrea Agostinelli, Timo~I Denk, Zal{\'a}n Borsos, Jesse Engel, Mauro Verzetti,
  Antoine Caillon, Qingqing Huang, Aren Jansen, Adam Roberts, Marco
  Tagliasacchi, et~al.
\newblock Musiclm: Generating music from text.
\newblock {\em arXiv preprint arXiv:2301.11325}, 2023.

\bibitem{singer2022make}
Uriel Singer, Adam Polyak, Thomas Hayes, Xi~Yin, Jie An, Songyang Zhang, Qiyuan
  Hu, Harry Yang, Oron Ashual, Oran Gafni, et~al.
\newblock Make-a-video: Text-to-video generation without text-video data.
\newblock {\em arXiv preprint arXiv:2209.14792}, 2022.

\bibitem{mcdonald2022great}
Joseph McDonald, Baolin Li, Nathan Frey, Devesh Tiwari, Vijay Gadepally, and
  Siddharth Samsi.
\newblock Great power, great responsibility: Recommendations for reducing
  energy for training language models.
\newblock In {\em Findings of the Association for Computational Linguistics:
  NAACL 2022}, pages 1962--1970, 2022.

\bibitem{openai}
OpenAI.
\newblock Transforming work and creativity with ai.
\newblock \url{https://openai.com/product}.
\newblock Accessed: 2023-03-23.

\bibitem{forefront_ai}
Forefront.
\newblock Powerful language models a click away.
\newblock \url{https://forefront.ai/}.
\newblock Accessed: 2023-03-23.

\bibitem{ai21}
AI21 Labs.
\newblock When machines become thought partners.
\newblock \url{https://ai21.com/}.
\newblock Accessed: 2023-03-23.

\bibitem{anand2022trust}
SVR Anand, Serhat Arslan, Rajat Chopra, Sachin Katti, Milind~Kumar Vaddiraju,
  Ranvir Rana, Peiyao Sheng, Himanshu Tyagi, and Pramod Viswanath.
\newblock Trust-free service measurement and payments for decentralized
  cellular networks.
\newblock In {\em Proceedings of the 21st ACM Workshop on Hot Topics in
  Networks}, pages 68--75, 2022.

\bibitem{witnesschain}
Witness~Chain team.
\newblock Witness chain.
\newblock \url{https://www.witnesschain.com/}.
\newblock Accessed: 2023-07-16.

\bibitem{eigenlayer}
Eigenlayer.
\newblock \url{https://www.eigenlayer.xyz/}.
\newblock Accessed: 2023-07-17.

\bibitem{sla4openAPI}
Sla4oai-specification.
\newblock \url{ https://github.com/isa-group/SLA4OAI-Specification}, 2022.

\bibitem{kang2022scaling}
Daniel Kang, Tatsunori Hashimoto, Ion Stoica, and Yi~Sun.
\newblock Scaling up trustless dnn inference with zero-knowledge proofs.
\newblock {\em arXiv preprint arXiv:2210.08674}, 2022.

\bibitem{kalodner2018arbitrum}
Harry Kalodner, Steven Goldfeder, Xiaoqi Chen, S~Matthew Weinberg, and Edward~W
  Felten.
\newblock Arbitrum: Scalable, private smart contracts.
\newblock In {\em 27th $\{$USENIX$\}$ Security Symposium ($\{$USENIX$\}$
  Security 18)}, pages 1353--1370, 2018.

\bibitem{szegedy2015going}
Christian Szegedy, Wei Liu, Yangqing Jia, Pierre Sermanet, Scott Reed, Dragomir
  Anguelov, Dumitru Erhan, Vincent Vanhoucke, and Andrew Rabinovich.
\newblock Going deeper with convolutions.
\newblock In {\em Proceedings of the IEEE conference on computer vision and
  pattern recognition}, pages 1--9, 2015.

\bibitem{uchida2017embedding}
Yusuke Uchida, Yuki Nagai, Shigeyuki Sakazawa, and Shin'ichi Satoh.
\newblock Embedding watermarks into deep neural networks.
\newblock In {\em Proceedings of the 2017 ACM on international conference on
  multimedia retrieval}, pages 269--277, 2017.

\bibitem{darvish2019deepsigns}
Bita Darvish~Rouhani, Huili Chen, and Farinaz Koushanfar.
\newblock Deepsigns: An end-to-end watermarking framework for ownership
  protection of deep neural networks.
\newblock In {\em Proceedings of the Twenty-Fourth International Conference on
  Architectural Support for Programming Languages and Operating Systems}, pages
  485--497, 2019.

\bibitem{fan2019rethinking}
Lixin Fan, Kam~Woh Ng, and Chee~Seng Chan.
\newblock Rethinking deep neural network ownership verification: Embedding
  passports to defeat ambiguity attacks.
\newblock {\em Advances in neural information processing systems}, 32, 2019.

\bibitem{adi2018turning}
Yossi Adi, Carsten Baum, Moustapha Cisse, Benny Pinkas, and Joseph Keshet.
\newblock Turning your weakness into a strength: Watermarking deep neural
  networks by backdooring.
\newblock In {\em 27th USENIX Security Symposium (USENIX Security 18)}, pages
  1615--1631, 2018.

\bibitem{szyller2021dawn}
Sebastian Szyller, Buse~Gul Atli, Samuel Marchal, and N~Asokan.
\newblock Dawn: Dynamic adversarial watermarking of neural networks.
\newblock In {\em Proceedings of the 29th ACM International Conference on
  Multimedia}, pages 4417--4425, 2021.

\bibitem{fernandez2023stable}
Pierre Fernandez, Guillaume Couairon, Herv{\'e} J{\'e}gou, Matthijs Douze, and
  Teddy Furon.
\newblock The stable signature: Rooting watermarks in latent diffusion models.
\newblock {\em arXiv preprint arXiv:2303.15435}, 2023.

\bibitem{zhao2023recipe}
Yunqing Zhao, Tianyu Pang, Chao Du, Xiao Yang, Ngai-Man Cheung, and Min Lin.
\newblock A recipe for watermarking diffusion models.
\newblock {\em arXiv preprint arXiv:2303.10137}, 2023.

\bibitem{christ2023undetectable}
Miranda Christ, Sam Gunn, and Or~Zamir.
\newblock Undetectable watermarks for language models.
\newblock {\em arXiv preprint arXiv:2306.09194}, 2023.

\bibitem{kirchenbauer2023watermark}
John Kirchenbauer, Jonas Geiping, Yuxin Wen, Jonathan Katz, Ian Miers, and Tom
  Goldstein.
\newblock A watermark for large language models.
\newblock {\em arXiv preprint arXiv:2301.10226}, 2023.

\bibitem{lukas2022sok}
Nils Lukas, Edward Jiang, Xinda Li, and Florian Kerschbaum.
\newblock Sok: How robust is image classification deep neural network
  watermarking?
\newblock In {\em 2022 IEEE Symposium on Security and Privacy (SP)}, pages
  787--804. IEEE, 2022.

\bibitem{yan2023rethinking}
Yifan Yan, Xudong Pan, Mi~Zhang, and Min Yang.
\newblock Rethinking white-box watermarks on deep learning models under neural
  structural obfuscation.
\newblock In {\em 32th USENIX security symposium (USENIX Security 23)}, 2023.

\bibitem{liu2023false}
Jian Liu, Rui Zhang, Sebastian Szyller, Kui Ren, and N~Asokan.
\newblock False claims against model ownership resolution.
\newblock {\em arXiv preprint arXiv:2304.06607}, 2023.

\end{thebibliography}
\bibliographystyle{unsrt}
\end{document}